# High-precision Absolute Distance Measurements over a Long Range Based on Two Optoelectronic Oscillators

Jinlong Yu[(1)], Ju Wang[(1)], Wang Miao[(1)], Jigui Zhu[(2)], Bin Sun[(1)], Wenrui Wang[(1)] and Hao Hu[(3)]

(1) Laboratory of Optical Fiber Communication, School of Electrical and Information Engineering, Tianjin University, Tianjin 300072, China

(2) State Key Laboratory of Precision Measuring Technology and Instruments, Tianjin University, Tianjin 300072, China

(3) DTU Fotonik, Technical University of Denmark, Ørsteds Plads, Building 343, DK-2800 Kgs. Lyngby, Denmark

Email: yujinlong@tju.edu.cn; wangju@tju.edu.cn

**Abstract:**

**Absolute distance measurement (ADM) over a long range has been studied intensely over the last several decades, due to its important applications in large-scale manufacturing and outer space explorations [1-5]. Traditional absolute distance measurements utilize detection of time-of-flight information, detection of phase shift, or a combination of the two [6-17]. In this paper, we present a novel scheme for high-precision ADM over a long range based on frequency detection by using two optoelectronic oscillators (OEO) to convert distance information to frequency information. By taking advantage of accumulative magnification theory, the absolute error of the measured distance is magnified by about $2\times10^5$ times, which makes the precision of the measured distance significantly improved. In our experiments, the maximum error is 1.5 μm at the emulated ~6 km distance, including the drift error of about 1 μm in the air path due to the change in environmental conditions. In addition, the measurable distance using this scheme could be further extended. The highest relative measurement precision is $2\times10^{10}$ in our current system while the actual relative measurement**

**precision of our experimental system is limited by the variation of atmospheric conditions and is about $4\times10^9$.**

**Introduction:**

Length is one of the seven fundamental physical quantities. The accuracy of distance measurement progresses with human understanding of science and in turn contributes to technology advancement. High-precision absolute distance measurement (ADM) over a long range is of fundamental importance for applications such as manufacturing of large-scale structures, systems with large geometric scales, and outer space explorations [1-5]. The rapid development of tight formation flying satellite missions in recent years places a higher than ever requirement on ADM precision and range, which stimulates significant research interest in the field [2].

Traditional distance measurements often take advantage of two general methods. The first method is based on time-of-flight by directly measuring the round trip time delay of a pulse. The time-of-flight methods have shown inadequate resolution due to the limited time measurement capability of current optical/electrical instruments [6-8]. The second method is based on the detection of phase shift of a signal after traveling a given distance, which offers good precision but has the $2\pi$ phase ambiguity issue and is only suitable for measurement of incremental displacement [6, 9-11]. A multi-wavelength optical interferometer can somewhat extend the ambiguity range and achieve good resolution [12-15]. However it requires either a tunable laser source or multiple laser sources and the accuracy of ADM is limited by the wavelength precision and stability of the individual laser used [16, 17]. Recently the availability of femtosecond optical frequency combs has opened new perspectives in the field of high-precision ADM over a long range [18-25]. However, this method highly relies on either stability of the pulse repetition rate or measurement precision of the phase of the pulse-to-pulse carrier envelope.

In general, accurate distance measurement requires a precise ruler and a proper method to compare the distance to be measured with the ruler. In an optoelectronic oscillator (OEO), there is one-to-one correspondence between the length of the loop (the measured distance can be a portion of the loop) and the resonant fundamental frequency. A key feature of the OEO configuration is that it usually oscillates at a high-order harmonic. And because of the low loss characteristics of fiber, the output microwave circuit has high Q value, which means that the measurement of the oscillation frequency could be of high precision. The high-order resonant frequencies are many times higher than the fundamental frequency. As a result, by measuring the high order resonant frequency accurately, the error of the fundamental frequency could be reduced many times. Correspondingly, the accuracy of loop length (distance measurement) is significantly improved. Building upon this concept, we developed and demonstrated a novel OEO scheme for high-precision ADM over a long range.

**Measurement Concept**

In 1995, the concept of optoelectronic oscillator (OEO) was introduced by X. S. Yao and L. Maleki [26-27]. As shown in figure 1, a typical OEO includes an optical intensity modulator, optical fiber, photodetector, microwave amplifier and filter, all of which compose a feedback loop. A distance path defined by reflection of a target is involved as one portion of the OEO's loop.

The fundamental frequency of the OEO is determined by the total group delay of the loop, including the physical length delay of the loop and the group delay resulting from dispersive components in the loop. The fundamental resonant frequency is given by $f_b=1/\tau$, where $\tau$ is the total group delay which consists of two parts: the reference delay $\tau_0$ and the additional delay $\tau_L$. The reference delay time $\tau_0$ includes the group delay of the electrical devices and fiber. The additional delay is related to the distance to be measured by $\tau_L=2Ln/c$, where $n$ denotes the effective refractive index at the measured atmospheric conditions and $c$ is the speed of light in vacuum. The distance $L$ can thus be calculated using $f_b$ in the following equation:

$$L = \frac{1}{2}\left(\frac{c}{nf_b} - \frac{c\tau_0}{n}\right) \tag{1}$$

A microwave filter is placed in the loop to select a high-order resonant harmonics mode to oscillate, e.g. the $N$-th order mode where $N$ is a very large integer number. The oscillation frequency of the $N$-th mode $f_N$ is related to the fundamental frequency by [26]

$$f_N = Nf_b \tag{2}$$

The measured distance $L$ can also be calculated using $f_N$ and $N$ in the following formula:

$$L = \frac{1}{2}\left(\frac{Nc}{nf_N} - \frac{c\tau_0}{n}\right) \tag{3}$$

The fundamental frequency of the OEO is in the kilohertz range in correspondence with a fiber length of over 1 km. Most OEOs can generate microwave resonant signals of high-order harmonics at several or tens of gigahertz which is 4~6 orders of magnitude higher than the fundamental frequency. If the measured distance is calculated from $f_N$ and $N$, the measured error would be reduced with the factor $1/N$ in comparison with detecting the fundamental frequency directly under the premise of the same accuracy of the frequency measurement instrument. For example, the fundamental frequency of an OEO with a loop of 500-m distance is approximately 300 kHz. A 50 nm length change in a 500 m distance would produce $3\times10^{-5}$ Hz frequency shift at the fundamental frequency. However, if the OEO operates at a high-order harmonic frequency of 30 GHz (i.e. $N=10^5$), a frequency shift of 3 Hz can be obtained. In other words, the accuracy of 50 nm could be achieved in 500-m range so long as the measurement accuracy of the ~30 GHz frequency is better than 3 Hz, which would be measured much more easily than that of measuring $3\times10^{-5}$ Hz shift @300 kHz using the same instrument.

The derivative of $L$ with respect to $f_N$ is

$$\Delta L = -\frac{cN}{2f_N^2 n}\Delta f_N = -\frac{cN}{2f_N n}\frac{1}{f_N}\Delta f_N = -(L+\frac{c\tau_0}{2n})\frac{1}{f_N}\Delta f_N \tag{4}$$

Assuming that the uncertainty of frequency measurement is $\Delta f_N$, which is commonly determined by the accuracy of the measurement instrument, the relative precision of the distance measurement can be defined as:

$$\left|\frac{L}{\Delta L}\right| = \left(\frac{2nL}{2nL+c\tau_0}\right)\left|\frac{f_N}{\Delta f_N}\right| \tag{5}$$

As clearly indicated in Equation (5), the measurement relative precision is proportional to the uncertainty of the high-order resonant frequency.

In addition to the resonant frequency $f_N$, using Equation (3) to calculate the distance $L$ requires knowing the order of the resonant mode $N$. The value of $N$ can be obtained using the following equation:

$$N = \left\lfloor \frac{f_N}{f_{b-rough}} \right\rfloor \tag{6}$$

where, $\lfloor\ \rfloor$ denotes rounding of a number to the nearest integer and $f_{b-rough}$ is the roughly measured fundamental frequency as the frequency spacing between two adjacent resonant peaks. The target absolute distance $L$ can be worked out as:

$$L = \frac{1}{2}\left(\frac{\left\lfloor \frac{f_N}{f_{b-rough}} \right\rfloor c}{n f_N} - \frac{c\tau_0}{n}\right) \quad (7)$$

Assuming that the instrument has an uncertainty of $\Delta f_N$ in frequency measurement, the uncertainty of $N$ is

$$\Delta N = \frac{\Delta f_N}{f_{b-rough}} - \frac{f_N}{f_{b-rough}^2}\Delta f_{b-rough} \approx -\frac{f_N}{f_{b-rough}^2}\Delta f_{b-rough} = -N\frac{\Delta f_{b-rough}}{f_{b-rough}} \quad (8)$$

The typical fundamental frequency for the measured distance of one to ten kilometers is in the range 15~150 kHz. The typical value of N is of the order of $10^5$. If we can control the uncertainty of frequency measurement $\left|\frac{\Delta f_{b-rough}}{f_{b-rough}}\right|$ to be smaller than $5\times10^{-6}$, the accuracy of $N$ is ensured according to Equation (8).

**Experimental setup**

Figure 2 shows the block diagram of the experimental setup for verification of the OEO based absolute distance measurement. An optical dual-loop OEO with orthogonal polarizations is adopted in order to obtain a high side mode suppression ratio [28, 29].

Two distributed feedback lasers (LD1 and LD2) are used as the light sources with their wavelengths of $\lambda_1$ and $\lambda_2$, respectively. The two laser beams are combined by a wavelength-division multiplexer (WDM1) and then injected into a Mach-Zehnder type LiNbO$_3$ optical intensity modulator. An Erbium-doped fiber amplifier (EDFA) is used after the modulator to compensate the optical loss of the system.

After the EDFA, $\lambda_1$ is separated by an optical add-drop multiplexer (OADM1) from $\lambda_2$ and goes into the port 1 of a circulator. Exiting from port 2 of the circulator, $\lambda_1$ is beam expanded and then propagates along the spatial distance to be measured to reach the target. The target is movable so that the distance can be varied to test the system. On the other side of the target, a Michelson interferometer is implemented so that the displacement of the target can be precisely measured to validate the system performance. The reflected beam from the target is recoupled back to port 2, routed to port 3 of the circulator and combined with $\lambda_2$ through another optical add-drop multiplexer (OADM2). The combined light is then fed into the OEO oscillator.

In a typical OEO system, the length of the loop is susceptible to environmental influences such as temperature, mechanical vibration and so on. Therefore stabilization of the loop is particularly important to achieve the desired measurement accuracy. In the design, the two laser beams pass through different optical paths. As a result, two OEOs are established. The optical path of $\lambda_1$ includes the distance to be measured while that of $\lambda_2$ does not. The first dual-loop OEO with orthogonal polarizations where $\lambda_1$ propagates is used to measure the distance and is defined as measurement oscillator. The second dual-loop OEO where $\lambda_2$ propagates serves to stabilize the reference delay length and is defined as stabilization oscillator. The two oscillator frequencies are defined as $f_N$ and $f_R$, respectively. In addition, the two-oscillator design offers the unique feature of self-referencing to minimize the environmental influences.

A long section of single mode fiber (SMF) is placed in the optical portion of the OEO to emulate a long distance in space. Because the length of the SMF drift is about 1μm/℃/m [30], the long SMF is placed in the common loop of the measurement and stabilization oscillators so that the fiber length drift induced errors can be minimized. Since the spectral purity of OEO will be better for longer distance to be measured [26], the measurable distance using this scheme could be further extended. For a short distance to be measured, a long SMF should be included in the common loop in order to enhance the spectral purity of OEO.

After transmitting through the SMF, the light is split into two orthogonal polarization paths through a polarization-beam splitter (PBS). Both paths use polarization maintaining fibers (PMF) to preserve the polarization states. One path (loop ① in Figure 2) has a PMF length that is 120 m longer than the other one. An optical switch is also placed in loop ① for the purpose of calculating the value of $N$ (discussed later). The two paths are then combined through a polarization-beam combiner (PBC) and fed into the microwave portion to complete the OEO operation. The two loops have different loop lengths which can significantly suppress the side mode. The entire optical portion of the OEO is placed inside a temperature controlled box with temperature variation less than 0.1°C.

Exiting from the optical dual-loop, the laser light enters a 100-m PMF coiled on a piezoelectric transducer (PZT) which controls the loop length and stabilizes the oscillator. Then the two wavelengths are separated by another wavelength-division multiplexer (WDM2) and detected by two photodetectors PD1 and PD2, respectively. The converted electrical signals pass through two RF bandpass filters BPF1 centered at 20 GHz and BPF2 centered at 5 GHz. The bandpass filters are used to achieve single mode oscillation at high frequencies. The band pass filtered outputs are amplified separately and then combined together using an RF combiner to drive the LiNbO$_3$ modulator.

In the measurement oscillator formed by λ1, there are two RF splitters (RF S1 and RF S2) to tap out signals for different purposes. At the output of RF S1, an envelope detector and a frequency counter (Gwinstek GFC-8010H, 1 miliHz accuracy) are accessed in sequence to measure the fundamental frequency of the measurement oscillator roughly and further to obtain the value of *N*. At the output of RF S2, the high-order mode resonant frequency $f_N$ of the measurement oscillator is detected by an RF spectrum analyzer (Agilent 8564E).

In the reference oscillator formed by λ₂, an RF splitter (RF S3) is inserted to tap out a small amount of microwave power to stabilize the common optical path of the oscillators though a phase-locked loop (PLL) and the PZT. The high-order mode oscillation frequency from the stabilization oscillator is phase-locked to a local frequency from a crystal oscillator. The PLL controls PZT to compensate the change in loop length caused by environmental influences. In the detailed implementation, the oscillation frequency of the stabilization oscillator $f_R$ is first divided by 500 using an adjustable divider and then mixed with a local reference frequency $f_{ref}$ produced by a highly stabile 10 MHz crystal oscillator with frequency stability better than $10^{-11}$ in a double-balanced mixer. The mixed signal passes through a low-pass filter, feeds into a high-voltage converter, and drives the PZT-based fiber stretcher. The change of the driving voltage applied to PZT will lead to expansion or contraction of the fiber stretcher, and causes a small change in the optical loop length until the difference in frequencies becomes zero. In this way the oscillation frequency of the stabilization OEO is locked to a certain value around 5 GHz (the center frequency of BPF2) with an actual measured uncertainty of $2.5 \times 10^{-10}$. The feedback control mechanism ensures that the optical loop length is maintained constant.

Note that, the measurement oscillator and the stabilization oscillator share most of the OEO path except for the distance under test and the microwave sections. Therefore, the referencing mechanism could compensate most of the loop length drifts of the oscillator (e.g. the long SMF). The residual error caused by the difference between the two individual microwave sections can be further minimized by placing them closely together in a temperature controlled (~0.1°C variation) chamber. By doing so, the difference between the two microwave sections is not taken into consideration in the current system.

**Results and Discussions**

As shown in Figure 3 (a), the designed OEO produced a high spectral purity oscillation frequency at about 20 GHz in the measurement oscillator with the long SMF of about 4 km. The sidemode suppression ratio reached 60 dB by utilizing the dual-loop design with orthogonal polarizations. The oscillation frequency $f_N$ =19992419218 Hz, has a full width at half-maximum $\Delta f_{FWHM}$=1 Hz when measured by the electrical spectrum analyzer (Agilent 8564E) with a resolution bandwidth (RBW) of 1 Hz and a span of 100 Hz. According to Equation (5), the highest relative precision of the distance measurement is $2\times10^{10}$ in our current system. On the other hand as shown in Figure 3 (b), when the system operated in a single loop configuration where path ① was switched off, the sidemode suppression ratio was only 20 dB and there were a number of side modes in the bandwidth (30 MHz) of RF bandpass filter. Envelope of the resonant microwave is detected by the envelope detector. The following frequency counter detected the fundamental frequency roughly as the frequency spacing between two adjacent resonant peaks, $f_{b\text{-}rough}$=44728.972Hz , with an accuracy of 1 mHz. The relative uncertainty of frequency measurement, $\left|\dfrac{\Delta f_{b\text{-}rough}}{f_{b\text{-}rough}}\right| = 2.2\times10^{-8}$, is less than $5\times10^{-6}$, and ensure the correct value of $N$, which is 446968 according to Equation (6). The measured distance $L$ can be calculated using Equation (7), $L$=3351211.124 mm, including the reference distance.

We conducted a series of experiments to demonstrate the capability of the OEO based distance measurement system by comparing the measurement results with the "truth" data. The "truth" data was provided by a standard fringe counting interferometric distance meter (Renishaw XL-80) [24].

In the first experiment, measurement of a 1.5-m distance in space was performed. The length of the SMF was about 4 km (equivalent to single-trip distance of about 3 km in space, the refractive index of the optical fiber $n_f$ =1.45 in the calculation, *(Length of SMF×Refractive index of SMF)/2)*. The distance to be measured was created by moving the target along a 1.5-m sideway at a step of 100 mm. The target had a plane mirror with a reflection coefficient of 95%. The atmospheric conditions (air temperature, pressure and humidity) were assumed to be constant with a refractive index of 1 for the entire measured distance of 1.5 m. In Figure 4 is plotted the comparison (left axis) between the distance measured by the OEO system and that measured using the Renishaw distance meter. The two sets of data matched very well. The differences between the two methods at the target positions are also plotted in the same figure (right axis). The error bars are the standard deviation of the mean value of 30 measurements. The mean values of differences were within ±1.4 μm while the standard deviations were all below 3.5 μm. We believe that the variation in air refractive index along the free space distance was a major contribution to the measurement uncertainty. According to reference [24], the variation in the atmospheric conditions could cause 1 μm variation in distance measurement. Another primary source of error was the length variation of the OEO loop, especially the long SMF line in the optical loop. For example, the optical length of the fiber can vary with environmental temperature fluctuations as a result of the thermo-optic effect and thermal expansion of the fiber. The phase-locked stabilized OEO and the actively temperature-controlled chamber can significantly reduce the environmental temperature fluctuation induced error. However, the residual error will still contribute to the measurement uncertainty.

To test the ranging capability of the measurement system, the length of the SMF was varied from approximately 1.5 km to 8 km, emulating the space distance variation from 1.1 km to 6 km (*(Length of SMF×Refractive index of SMF)/2)*. A series of measurements with different SMF lengths were performed and the results are shown in Figure 5 (1)-(8). At each SMF length, the target was moved from 0 to 25 mm at a step of 1.5 mm. The maximum difference between the mean value of the OEO system and the Renishaw distance meter was below 1.5 μm for all conditions. There was no obvious correlation between the difference and the increase in SMF length. We conclude that the long SMF has a negligible effect on the mean value of the measurement results. However, the uncertainty of the measurements increased slightly with the increase in SMF length. Nevertheless, even with a long SMF length of 8 km (or 6 km in free space), the standard deviations of 30 measurement data were less than 4μm. The system had a high precision with only 1.5μm error and with a relative precision of $4\times10^9$ over the distance of 6 km.

**Conclusion**

In summary, a novel OEO based approach was demonstrated for measurement of absolute distance. The OEO based system takes advantage of the accumulative magnification effect at high-order mode resonant frequencies to achieve high measurement accuracy. A dual-loop design with orthogonal polarizations was implemented to suppress the side mode resonances and obtain high quality resonance signals. Two optical wavelengths were multiplexed into the system, one for probing and the other for referencing. By actively controlling the loop length using the reference wavelength, the system stability was significantly improved. The system was evaluated for distance measurement at different lengths ranging from 1.1 km to 6 km. When compared with a commercial interferometric distance meter, the difference in mean value was smaller than 1.5 μm for all conditions and the difference did not increase when the distance increased. The maximum uncertainty was less than 4 μm (measured as the standard deviation taken over 30 measurements) and the uncertainty increased only slightly with the increase of distance. We estimated that the system had relative measurement accuracy better than $4\times10^9$, which also included the contribution of the variation of atmospheric conditions along the measurement path. The results illustrate the capability of the system for unambiguous measurement of distance over a large range, which may find important applications in future tight formation flying satellite missions.

**Acknowledgements**

The authors would like to thank Hai Xiao (Missouri University of Science and Tech.) and Palle Jeppesen (Technical University of Denmark) for the useful discussion and comments.

**Author contributions**

The project was planned by Jinlong Yu and Jigui Zhu. The experiments were designed by Jin long Yu and performed by Ju Wang, Wang Miao and Bin Sun. Wenrui Wang and Hao Hu assisted with the data analysis.

**Additional Information**

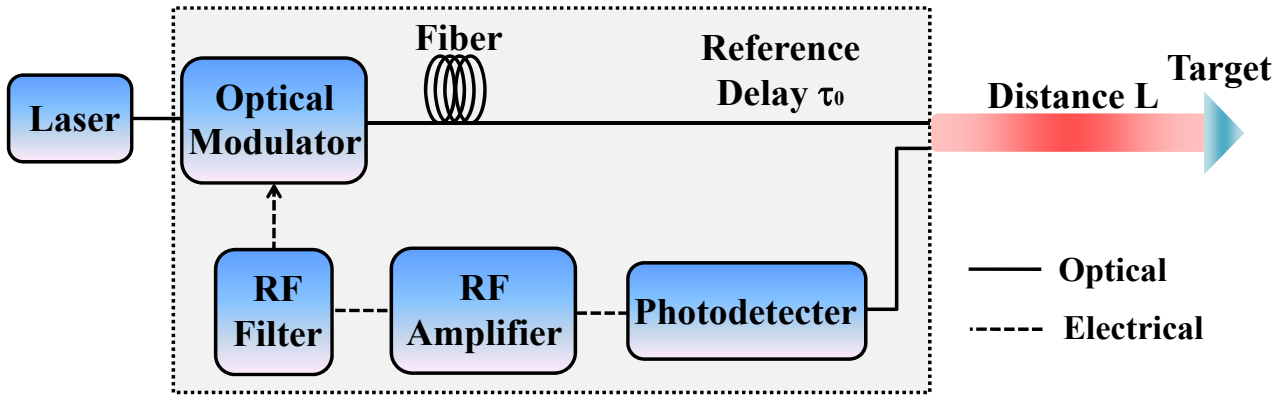

**Figure 1 | Basic schematic of optoelectronic oscillator based absolute distance measurement.**

The distance, which is defined by reflection off a target, is involved as one part of OEO's loop path.

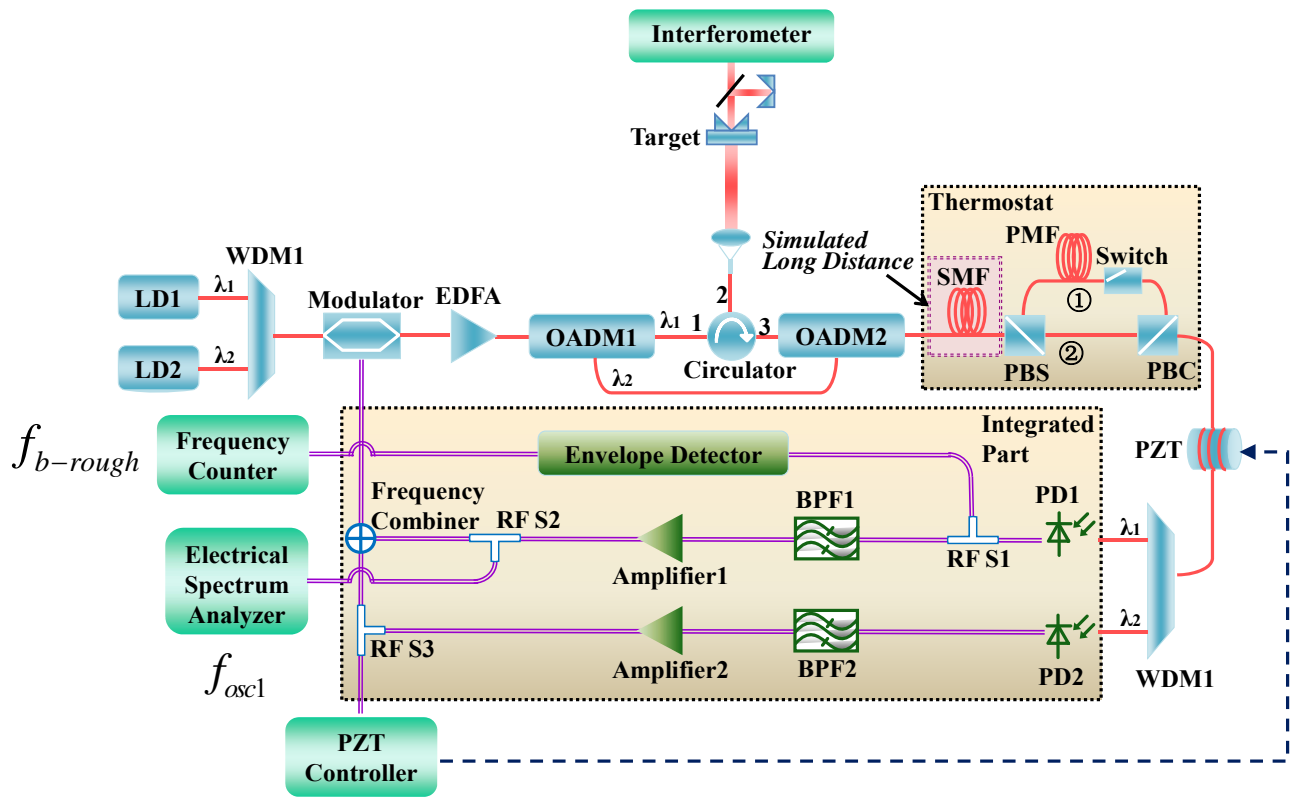

**Figure 2 | Block diagram of the experimental setup.** Each light beam ($\lambda_1$ or $\lambda_2$) propagates in a dual-loop OEO oscillator with orthogonal polarizations. The optical path of $\lambda_1$ includes the distance to be measured while that of $\lambda_2$ does not. The first dual-loop OEO where $\lambda_1$ propagates is used to

measure the distance, which is defined as measurement oscillator. The second dual-loop OEO where $\lambda_2$ propagates serves to stabilize the reference delay length and is defined as stabilization oscillator. A long single mode fiber (SMF) serves to emulate a long space distance and to improve spectral purity of OEO. The long SMF is placed in the common loop of the measurement and stabilization oscillators so that the fiber length drift induced errors can be minimized.

RF bandpass filters: BPF1 is centered at 20 GHz with a bandwidth of 30 MHz and BPF2 is centered at 5 GHz with a bandwidth of 30 MHz; RF S: RF Splitter ; WDM: wavelength-division multiplexer; EDFA: Erbium-doped fiber amplifier; OADM: optical add-drop multiplexer; PBS: polarization-beam splitter; PBC: polarization-beam combiner; PMF: polarization maintaining fiber; PZT: piezoelectric transducer; PLL: phase-locked loop.

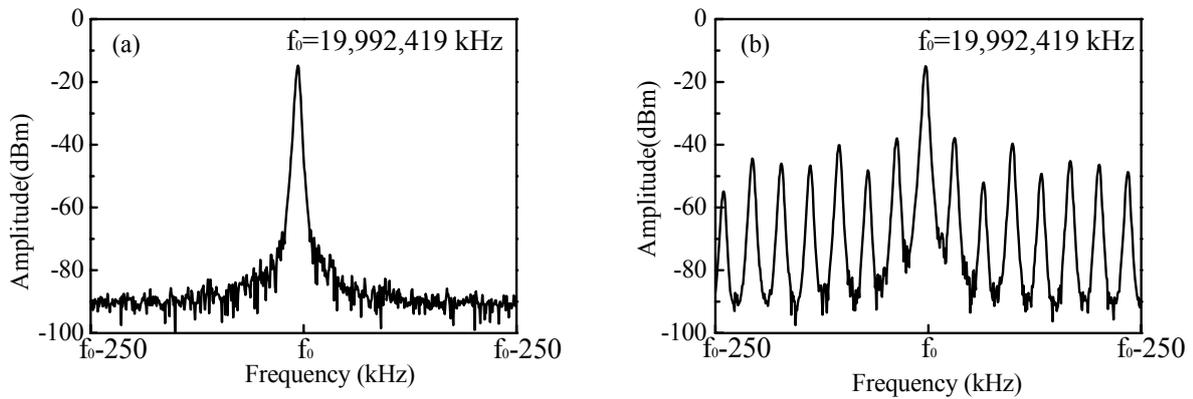

**Figure 3 | (a) RF spectrum of the measurement oscillator with dual-loop (SPAN 500 kHz, RBW 3 kHz). (b) RF spectrum of the measurement oscillator with single loop (the optical switch in path ① is off) (SPAN 500 kHz, RBW 3 kHz).**

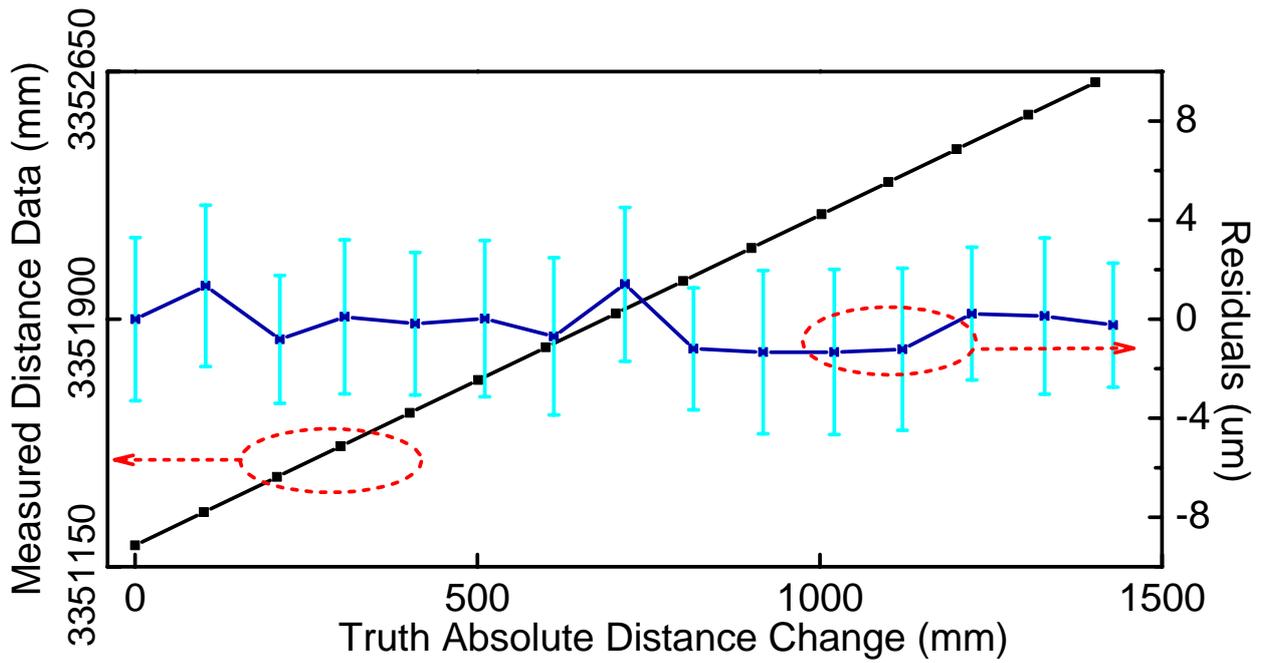

**Figure 4 | Measured distance and residuals of the measured distance over 1.5 m (at an emulated 3-km distance) versus truth data from a commercial Michelson interferometer.** Error bars are the standard deviation of the mean of 30 measurements. The errors included the contribution of the variation of atmospheric conditions along the measurement path.

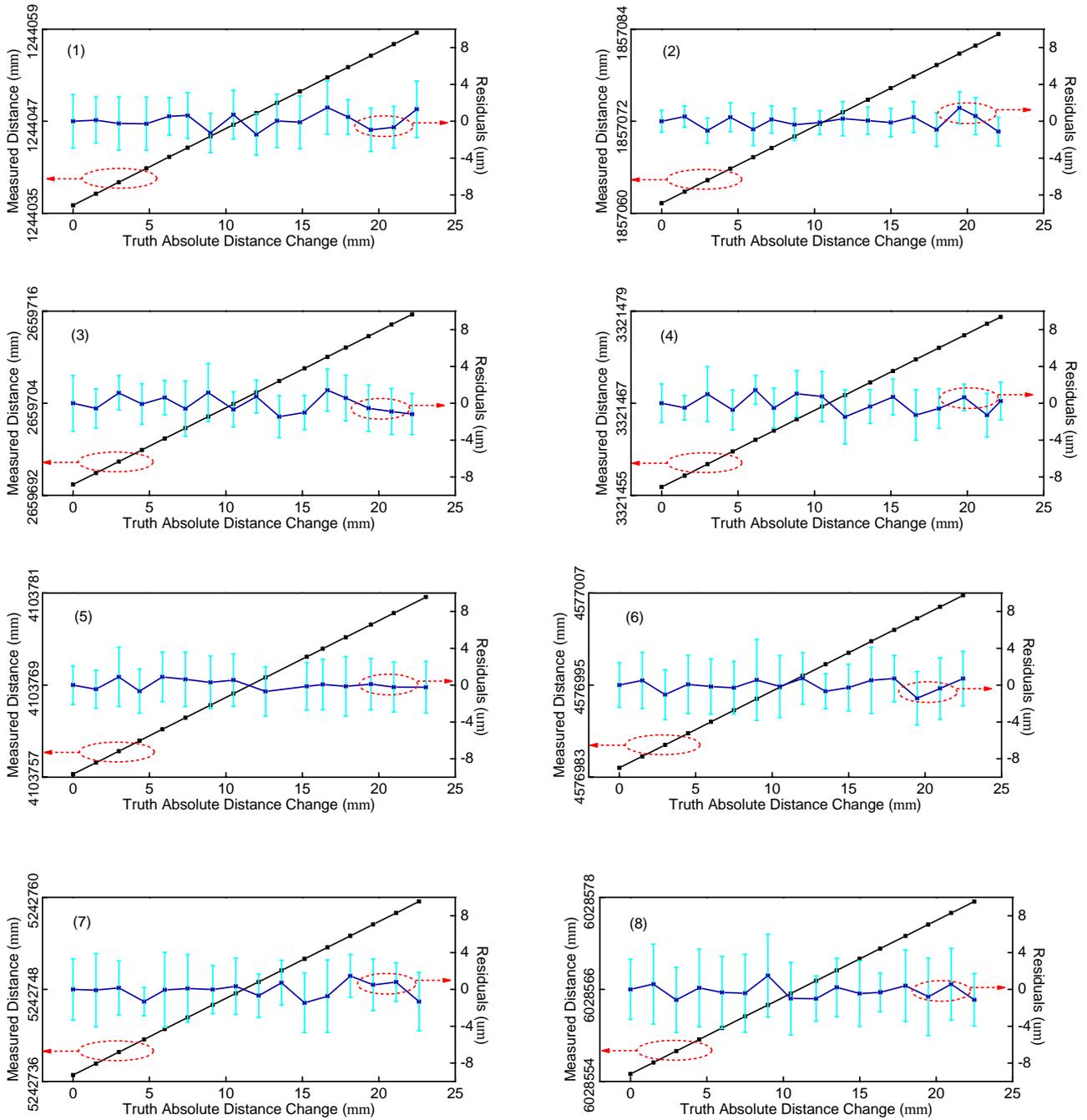

**Figure 5 | Ranging capability of our absolute distance measurement program.** The length of long SMF has been varied from about 1.5 km to 8 km in order to emulate the long space distance ranging from 1.1 km to 6 km. The target distance is updated with a step of 1.5 mm and varies from 0 to 25 mm. The measured distances and residuals of the measured distance for all the 8 cases are

shown in figure 5(1)-(8). There was no obvious correlation between the errors and the increase in SMF length.